\documentclass[11pt,twoside]{article}
\usepackage{asp2010}

\resetcounters

\begin{document}

\title{Signatures of an intermediate-age metal-rich bulge population}
\author{T.~Bensby,$^1$ S.~Feltzing,$^1$ A.~Gould,$^2$ J.A.~Johnson,$^2$ M.~Asplund,$^3$ D.~Ad\'en,$^1$ J.~Mel\'endez,$^4$ J.G.~Cohen,$^5$, I.~Thompson,$^6$, S.~Lucatello,$^7$ A.~Gal-Yam$^8$
\affil{
$^1$Lund Observatory, Box 43, SE-221\,00 Lund, Sweden\\
$^2$Department of Astronomy, Ohio State University, Columbus, OH, USA\\
$^3$Research School of Astronomy and Astrophysics, ANU, Weston,  Australia\\
$^4$Dpto de Astronomia do IAG/USP, Univ. de S\~ao Paulo, S\~ao Paulo, Brasil\\
$^5$Palomar Observatory, California Inst. of Technology, Pasadena, CA, USA\\
$^6$Carnegie Observatories, 813 Santa Barbara St., Pasadena, CA, USA\\
$^7$INAF-Astronomical Observatory of Padova, Padova, Italy\\
$^8$Benoziyo Center for Astrophysics, Weizmann Inst. of Science, Rehovot, Israel}}

\begin{abstract}
We have determined detailed elemental abundances and stellar ages 
 for a sample of now 38 microlensed dwarf and subgiant stars in the Galactic 
 bulge. Stars with sub-solar metallicities
 are all old and have enhanced $\alpha$-element abundances -- very similar to
 what is seen for local thick disk stars. The metal-rich stars
 on the other hand show a wide variety of stellar ages, ranging from 
 3-4\,Gyr to 12\,Gyr, and an average around 7-8\,Gyr. The existence of young and metal-rich 
 stars are in conflict with recent photometric studies of the bulge which
 claim that the bulge only contains old stars.
\end{abstract}

\section{Signatures on an intemediate-age metal-rich bulge population}

The first 26 microlensed dwarf and subgiant stars in the bulge 
from \cite{bensby2010,bensby2011} showed that the metallicity distribution 
of the Galactic bulge is likely bimodal with a paucity of stars around solar 
metallicities. Adding another 12 microlensed bulge dwarfs from the 2011 
observing campaign the metallicity distribution of the current sample 
of 38 microlensed dwarfs is still bi-modal with two distinct peaks: 
one metal-poor peak with 16 stars and an average metallicity of 
$\rm [Fe/H]\approx-0.6$ and one metal-rich peak with 22 stars and an 
average metallicity of $\rm [Fe/H]\approx+0.3$. A two-sided KS-test with 
the red giant sample in BaadeÕs window from \cite{zoccali2008}, re-analysed 
by \cite{hill2011}, gives a $p$-value of 0.47. This means that we can not 
reject the null hypothesis that the microlensed sample and the red giant 
sample are drawn from the same underlying metallicity distribution.

\begin{figure}
\resizebox{\hsize}{!}{
\includegraphics[bb=150 162 590 525,clip]{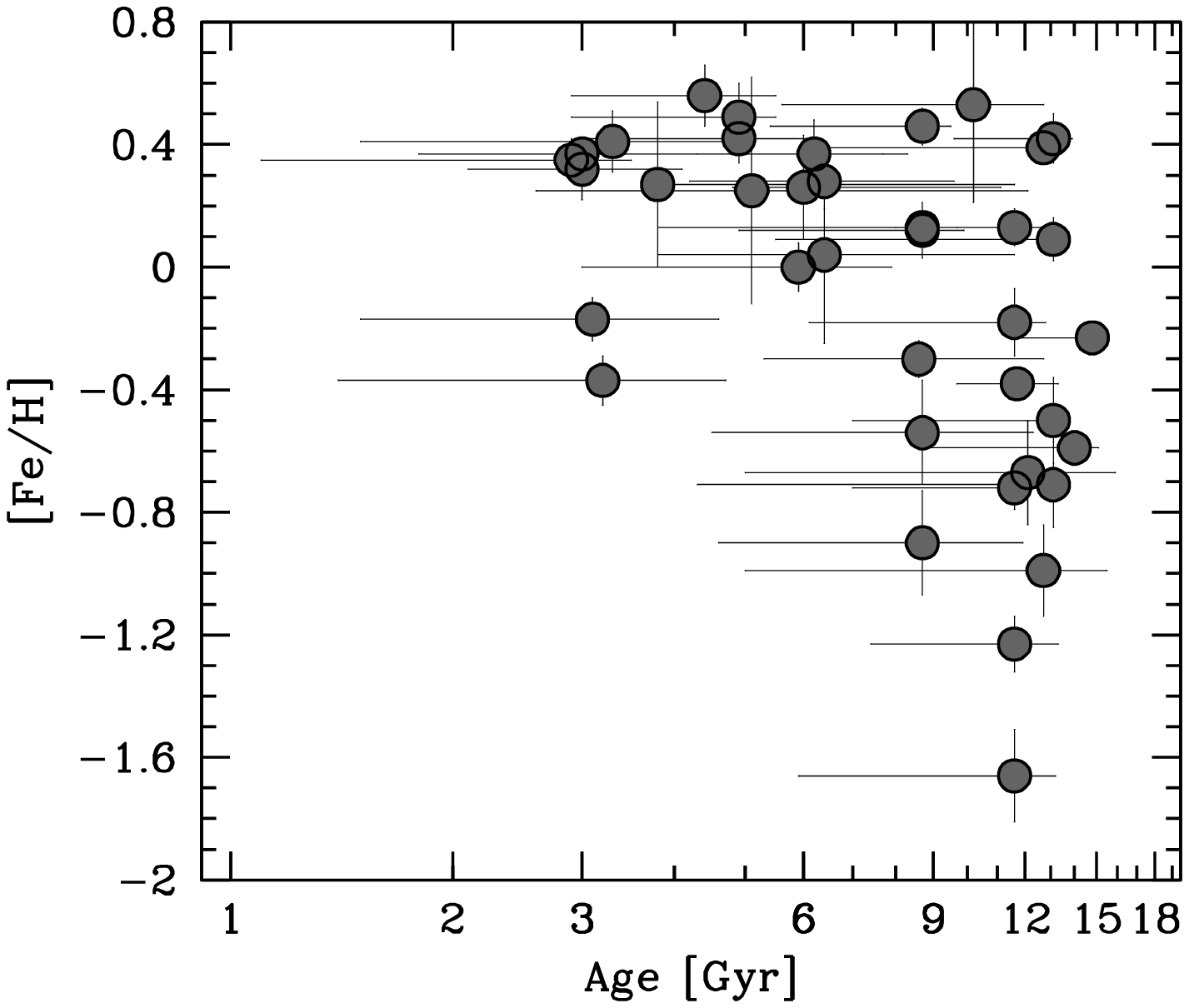}
\includegraphics[bb=20 162 580 525,clip]{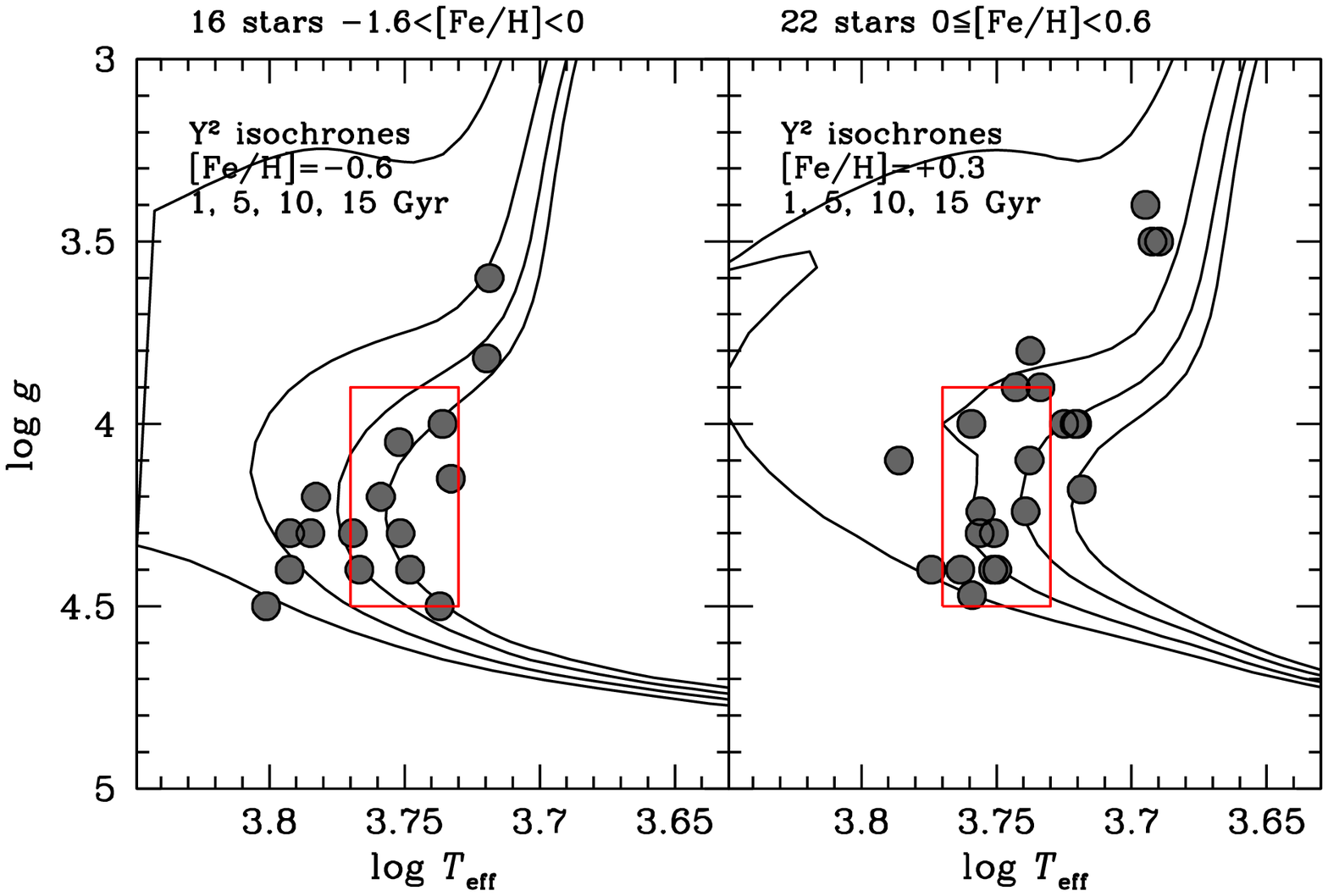}
}
\caption{{\sl Left:} Age-metallicity
diagram. {\sl Right:} The metal-poor and metal-rich bulge dwarfs plotted on isochrones
reprenentative of their respective average metallicity. The red rectangle
occupy the same region in each plot.
\label{fig:ages}
                    }
   \end{figure}

Figure~\ref{fig:ages} (left-hand plot) shows the age-metallicity diagram for 
the microlensed bulge dwarfs. At sub-solar metallicities the stars are 
pre-dominantly old with ages between 9 and 13\,Gyr. The average age is 10.6\,Gyr with 
a spread of 3.3\,Gyr.
The 22 stars at super-solar metallicities on the other hand show a wide range 
of ages from only a few billion years old to as old as the Universe,
i.e. spanning the full range of ages from the Galactic disk to the halo. 
The average age is 6.9\,Gyr with a 
spread of 3.6 Gyr for the  stars at super-solar [Fe/H].

Photometric studies toward the Galactic bulge appear to indicate that the bulge 
population is all old and that there are no signs of a young or 
intermediate age population in the HR diagrams \citep[e.g.,][]{zoccali2003,clarkson2008}.
However, the metallicity distribution of the bulge spans a large range of 
metallicities \citep[e.g.,][]{fulbright2007}. In the right-hand plot of Fig.~\ref{fig:ages}
we show the microlensed sample, divided into 
the metal-poor and metal-rich sub-samples, on top are shown isochrones 
with different metallicities. The rectangle outlines the same 
$T_{\rm eff}-\log g$ region in each of the two plots. It is clear that, 
if a metal-poor isochrone would be plotted on top of the metal-rich stars, 
they would all appear old. This demonstrates the importance of taking 
isochrones with a suitable range of metallicities into account when estimating 
the age of the bulge from HR diagrams.

In summary, the sample of microlensed dwarf stars in the bulge
shows evidence for a bi-modal bulge population and that the metal-rich
population has a significant fraction of intermediate-age stars
\citep[but see][]{nataf2012}.
This will be further investigated in an upcoming paper.

\acknowledgements T.B. was funded by grant No. 621-2009-3911 from The 
Swedish Research Council. Work by A.G. was supported by NSF Grant AST\,0757888. 

\bibliographystyle{asp2010}
\bibliography{referenser}

\begin{thebibliography}{}
\expandafter\ifx\csname natexlab\endcsname\relax\def\natexlab#1{#1}\fi
\expandafter\ifx\csname url\endcsname\relax
  \def\url#1{\texttt{#1}}\fi
\expandafter\ifx\csname urlprefix\endcsname\relax\def\urlprefix{URL }\fi
\providecommand{\eprint}[2][]{\url{#2}}

\bibitem[{{Bensby} et~al.(2011){Bensby}, {Ad{\'e}n}, {Mel{\'e}ndez}, {Gould},
  {Feltzing}, {Asplund}, {Johnson}, {Lucatello}, {Yee}, {Ram{\'{\i}}rez},
  {Cohen}, {Thompson}, {Bond}, {Gal-Yam}, {Han}, {Sumi}, {Suzuki}, {Wada},
  {Miyake}, {Furusawa}, {Ohmori}, {Saito}, {Tristram}, \&
  {Bennett}}]{bensby2011}
{Bensby}, T., {Ad{\'e}n}, D., {Mel{\'e}ndez}, J., {Gould}, A., {Feltzing}, S.,
  {Asplund}, M., {Johnson}, J.~A., {Lucatello}, S., {Yee}, J.~C.,
  {Ram{\'{\i}}rez}, I., {Cohen}, J.~G., {Thompson}, I., {Bond}, I.~A.,
  {Gal-Yam}, A., {Han}, C., {Sumi}, T., {Suzuki}, D., {Wada}, K., {Miyake}, N.,
  {Furusawa}, K., {Ohmori}, K., {Saito}, T., {Tristram}, P., \& {Bennett}, D.
  2011, \aap, 533, A134

\bibitem[{{Bensby} et~al.(2010){Bensby}, {Feltzing}, {Johnson}, {Gould},
  {Ad{\'e}n}, M., {Mel\'endez}, {Gal-Yam}, {Lucatello}, {Sana}, {Sumi},
  {Miyake}, {Suzuki}, {Han}, {Bond}, \& {Udalski}}]{bensby2010}
{Bensby}, T., {Feltzing}, S., {Johnson}, J.~A., {Gould}, A., {Ad{\'e}n}, D.,
  M., A., {Mel\'endez}, J., {Gal-Yam}, A., {Lucatello}, S., {Sana}, H., {Sumi},
  T., {Miyake}, N., {Suzuki}, D., {Han}, C., {Bond}, I., \& {Udalski}, A. 2010,
  \aap, 512, A41

\bibitem[{{Clarkson} et~al.(2008){Clarkson}, {Sahu}, {Anderson}, {Smith},
  {Brown}, {Rich}, {Casertano}, {Bond}, {Livio}, {Minniti}, {Panagia},
  {Renzini}, {Valenti}, \& {Zoccali}}]{clarkson2008}
{Clarkson}, W., {Sahu}, K., {Anderson}, J., et al.
  2008, \apj, 684, 1110

\bibitem[{{Fulbright} et~al.(2007){Fulbright}, {McWilliam}, \&
  {Rich}}]{fulbright2007}
{Fulbright}, J.~P., {McWilliam}, A., \& {Rich}, R.~M. 2007, \apj, 661, 1152

\bibitem[{{Hill} et~al.(2011){Hill}, {Lecureur}, {G{\'o}mez}, {Zoccali},
  {Schultheis}, {Babusiaux}, {Royer}, {Barbuy}, {Arenou}, {Minniti}, \&
  {Ortolani}}]{hill2011}
{Hill}, V., {Lecureur}, A., {G{\'o}mez}, A., et al.
  2011, \aap, 534, A80

\bibitem[{{Nataf} \& {Gould}(2011)}]{nataf2012}
{Nataf}, D.~M., \& {Gould}, A.~P. 2011, arXiv:1112.1072 [astro-ph.GA]

\bibitem[{{Zoccali} et~al.(2008){Zoccali}, {Hill}, {Lecureur}, {Barbuy},
  {Renzini}, {Minniti}, {G{\'o}mez}, \& {Ortolani}}]{zoccali2008}
{Zoccali}, M., {Hill}, V., {Lecureur}, A., et al.
  2008, \aap, 486, 177

\bibitem[{{Zoccali} et~al.(2003){Zoccali}, {Renzini}, {Ortolani}, {Greggio},
  {Saviane}, {Cassisi}, {Rejkuba}, {Barbuy}, {Rich}, \& {Bica}}]{zoccali2003}
{Zoccali}, M., {Renzini}, A., {Ortolani}, S., et al.
  2003, \aap, 399, 931

\end{thebibliography}

\end{document}